# Title: Engineering robust strain transmission in van der Waals heterostructure devices

**Authors:** John Cenker[1,†,*,a], Jordan Fonseca[1,†,b], Mai Nguyen[1,†], Chaowei Hu[1], Daniel G. Chica[2], Takashi Taniguchi[3], Kenji Watanabe[4], Xiaoyang Zhu[2], Xavier Roy[2], Jiun-Haw Chu[1], Xiaodong Xu[1,5,*]

[1]Department of Physics, University of Washington, Seattle, Washington 98195, USA
[2]Department of Chemistry, Columbia University, New York, NY 10027 USA
[3]Research Center for Materials Nanoarchitectonics, National Institute for Materials Science, 1-1 Namiki, Tsukuba 305-0044, Japan
[4]Research Center for Electronic and Optical Materials, National Institute for Materials Science, 1-1 Namiki, Tsukuba 305-0044, Japan
[5]Department of Materials Science and Engineering, University of Washington, Seattle, Washington 98195, USA
Current Affiliation:  [a] Department of Physics, Columbia University, New York, NY, 10027, USA
[b] National Institute of Standards and Technology, Quantum Sensors Division, Boulder, CO 80305, USA

*Corresponding author's email: jc6267@columbia.edu, xuxd@uw.edu

**Abstract:** Atomically thin van der Waals materials provide a highly tunable platform for exploring emergent quantum phenomena in solid state systems. Due to their remarkable mechanical strength, one enticing tuning knob is strain. However, the weak strain transfer of graphite and hBN, which are standard components of high-qualityvdW devices, poses fundamental challenges for high-strain experiments. Here, we investigate strain transmission in less-explored orthorhombic crystals and find robust transmission up to several percent  at cryogenic temperatures. We further show that strain can be efficiently transferred through these crystals to other 2D materials in traditional heterostructure devices. Using this capability, we demonstrate in-situ strain and gate control of the optical properties of monolayer $WS_2$ utilizing the high-$\kappa$ dielectric insulator $Bi_2SeO_5$ as a substrate. These results enable the exploration of combined cryo-strain and gate tuning in a variety of layered systems such as moiré heterostructures, air-sensitive 2D magnets and superconductors, and any gated 2D device.

Keywords: cryo-strain tuning, 2D materials, van der Waals heterostructures, monolayer $WS_2$

The mechanical deformation of a crystal lattice by strain can have profound effects on the material's properties[1–4]. In bulk crystals, strain has been shown to be an effective tuning knob of several quantum phenomena including magnetic[5,6], superconducting[7,8], and topological states of matter[9,10]. The accessible strain range in standard experiments is often limited by the tensile strength of the crystal, which depends on the number of defects and imperfections where fracture occurs. Reducing the dimensionality of the crystal from the 3D bulk to the 2D limit reduces the total number of defects, making atomically thin van der Waals (vdW) crystals the strongest materials ever measured[11,12]. In addition to their exceptional strength, these materials and their heterostructures host fascinating 2D quantum phases including correlated insulating states[13–15], nematicity[16–19], frustrated magnetic orders[20–24], and integer and fractional quantum Hall effects[10,25–31]. The combination of unique mechanical durability and rich physics make 2D vdW materials a promising and unique platform for strain engineering.

A necessity for many of the quantum phases to emerge, however, is high device quality and the ability to control the carrier density and applied electric field via electrostatic gating. This is usually achieved via encapsulation[32] with hexagonal boron nitride (hBN) and/or graphite. Specifically, the hBN layers screen the charge traps and roughness of the silicon substrate[32,33], protect air-sensitive materials from the environment[34], and serve as the dielectric in gated devices, typically in conjunction with atomically-smooth graphite gates. However, hBN and graphite have weak interfacial and interlayer strain transmission[35–37], thereby posing fundamental challenges to the realization of high-quality 2D devices capable of sustaining large strains. To date, the maximum in-situ strain tuning of hBN-encapsulated samples is well below 1% despite extensive efforts to increase strain transmission by clamping the sample with evaporated gold layers[38].

In this work, we investigate the strain transmission of several materials with the goal of maximizing the in-situ strain tuning capabilities in vdW heterostructures with concurrent electrostatic gate control at cryogenic temperatures. We begin with the layered orthorhombic crystal CrSBr (see crystal structure in Fig. 1a), as the strain response is well calibrated (Methods). Figure 1b shows Raman scattering from the phonon peak centered around 345 cm$^{-1}$ of an unclamped thin bulk (~ 50 nm) CrSBr flake deposited on a polyimide substrate. Data at two selected piezo voltages, $V_s = 0$ V (blue) and 100 V (black) applied to the strain cell, are shown at a nominal sample temperature of ~ 5 K (see **Methods** and Supplementary Fig. S1). Despite the lack of a metal clamping layer, the phonon mode centered near 345 cm$^{-1}$ exhibits a large redshift of ~ 6.8 cm$^{-1}$ when $V_s$ is increased from 0 V to 100 V. Using the previously determined[39] strain shift rate of ~ 4.2 cm$^{-1}$/%, this corresponds to a uniaxial tensile strain of ~ 1.6 %. The full strain dependence of the Raman peak is shown in Fig. 1c, demonstrating a continuous redshift through the entire accessible voltage range. Fitting the peak position at each piezo voltage reveals a linear application of strain with negligible change to the phonon peak width (Fig. 1d). These results demonstrate a remarkably robust interlayer strain transmission in CrSBr, standing in stark contrast with many other common vdW materials, such as graphite, whose Raman shift under strain quickly vanishes in the few layer and bulk limit[35]. Moreover, we find that the unclamped CrSBr samples can survive to exceptionally high strains, with some samples sustaining ~ 3 % strain before slippage, indicating robust interfacial strain transfer between the polymer substrate and the flake (Supplementary Fig. S2).

The efficient interlayer strain transmission, endurance to extreme strains, and well calibrated strain response make CrSBr a convenient sensor of strain transmission in vdW heterostructures. Unlike graphene and transition metal dichalcogenides (TMDCs) which require mono- or few-layer flakes for sensing functionality, exfoliated CrSBr flakes of any thickness can be used, significantly simplifying sample fabrication. We first test the strain transfer efficiency of the ubiquitously used hexagonal boron nitride (hBN), which has a hexagonal lattice as shown in Figure 2a. In contrast to CrSBr, where the exterior Br bonds extend vertically along the stacking direction, the bonds in hBN are confined within the hexagonal plane (Fig. 2a). Moreover, the hexagonal crystal structure exhibits two stable stacking configurations separated by only a small energy barrier[40] (with AA′ being the most stable one), whereas CrSBr has only a single stable stacking configuration[39]. These differences in crystal structure may impact the tendency for soliton formation[41] in different crystals and contribute to the significant differences in interfacial and interlayer strain transmission.

Figure 2b shows the strain response of a CrSBr flake stacked on top of hBN (red) and another CrSBr flake deposited on the polyimide substrate adjacent to the hBN/CrSBr heterostructure (black). The effective strain applied to the CrSBr through the hBN is an order of magnitude less than that transferred to the CrSBr off the hBN. Moreover, the hBN sample exhibited extreme wrinkling after strain cycling (Supplementary Fig. S3), indicating that hBN is not suitable as a substrate for repeatable strain cycling at strain values above a fraction of a percent. These results demonstrate the poor strain transfer of hBN and confirm that the use of hBN can limit the strain tunability of vdW heterostructures and devices, consistent with the small strain ranges reported in previous measurements[38].

Since hBN is widely used as a substrate and gate dielectric in high-quality vdW devices, its poor strain transmission properties pose a significant challenge towards utilizing large strain as a tuning knob for the rich quantum phases which emerge in sufficiently clean 2D systems at cryogenic temperatures. The desirable mechanical properties of CrSBr, on the other hand, suggest that it may be used as an alternative component in strain-tuned vdW devices and heterostructures. However, the magnetic and semiconducting properties of CrSBr can result in significant magnetic proximity and charge transfer effects[42–45]. While these effects are interesting, an ideal straintronic dielectric substrate should provide an inert, atomically flat surface akin to hBN. So, the guiding principle is to explore layered materials with h-BN-like electronic properties and CrSBr-like structural properties, i.e., atomic bonds extending vertically out of the 2D basal plane.

The constantly expanding library of 2D materials provides new building blocks for devices with enhanced functionality. Two materials which have recently[46–50] been identified as promising dielectric insulators are $\alpha$-MoO$_3$ and Bi$_2$SeO$_5$. Unlike hBN and graphite, these crystals have orthorhombic crystal structures (Figs. 2c, e), with $\alpha$-MoO$_3$ in particular having a buckled, anisotropic crystal structure that resembles CrSBr. Figures 2d, f show the strain transmission characteristics of $\alpha$-MoO$_3$ and Bi$_2$SeO$_5$, respectively, as measured by a CrSBr strain sensor. The strain measured through both crystals is comparable with that of the CrSBr deposited directly on the polyimide substrate, demonstrating efficient strain transmission through the van der Waals bonded layers.

The exceptional stability and insulating nature[48] of Bi$_2$SeO$_5$ combined with its strain transfer properties make it a particularly appealing candidate for use as a hBN substitute in high-quality vdW devices. To explore this possibility, we fabricated a top-gated monolayer WS$_2$ device using Bi2SeO5 as a strain-transmitting, atomically flat dielectric substrate (see schematic in Fig.

3a and Methods). Figure 3b shows photoluminescence (PL) measurements of the monolayer $WS_2$ at select top gate voltages of −9.8 V and 7.3 V with the entire gate dependence shown in Supplementary Fig. S4. Due to the intrinsic n-type doping in monolayer $WS_2$, the negative gate voltage brings the monolayer closer to being charge neutral, while applying a positive top gate voltage efficiently dopes additional electrons into the monolayer.

We assign the three observed exciton features at low doping to be the neutral exciton ($X^0$), negatively charged trion ($X^-$), and biexciton (XX), based on a previous report[51]. In the highly doped regime, however, a single peak dominates the spectra. We label this peak $X^{-'}$, following the convention used for monolayer $WSe_2$. We note that the origin of $X^{-'}$ in TMD monolayers is still a subject of debate, with previous works in $WSe_2$ suggesting that this feature may arise from exotic many-body states[52], trion fine structures[53], or the second trion charging state[51,54]. The detailed experimental and theoretical understanding of this peak is beyond the scope of this work. Yet, the simplicity of the highly doped spectra makes it a more convenient strain probe than the undoped one which has several excitonic features. Importantly, the strain shift rate of $X^{-'}$ is very similar to the neutral exciton (see Figure 4c and Methods), enabling the use of either feature for strain calibration purposes and possibly offering an experimental observation that can be useful for future work to understand the true origin of this peak.

To check the device quality as well as the homogeneity of strain transmission in the heterostructure, we performed spatial mapping at different piezo and top gate voltages. Figure 3c shows the spatial dependence of PL from the sample with the gate voltage fixed at $V_g = 7.3$ V at zero piezo voltage. In this condition, strong PL is observed across the sample, with enhanced intensity at certain spots corresponding to bubbles or contamination commonly found in vdW heterostructures[55]. When the piezo voltage $V_s$ is increased, the spectra red shift as expected for tensile strain[56,57]. We extract the peak energy at every point with Gaussian fits and compare with the $V_s = 0$ V value (see Methods and Supplementary Fig. S5). This enables us to construct a strain map (Fig. 3d), i.e., the spatial dependence of strain at $V_s = 40$ V. The sample shows a large (> 1%), relatively homogeneous strain throughout the majority of the heterostructure region. These results confirm that $Bi_2SeO_5$ is capable of sustaining and transmitting large strains without compromising the high spectral quality of the encapsulated $WS_2$ monolayer. Therefore, these results demonstrate a new path for strain engineering in optoelectronic vdW devices that does not rely on metal clamping schemes, which are less robust to large strains and may also induce spatially inhomogeneous strain[38,58].

The ability to apply large strains to a monolayer semiconductor at cryogenic temperatures with concurrent gate control enables the study of strain effects on the rich excitonic species in these systems. Figures 4a and 4b show strain-dependent PL measurements taken at selected gate voltages of -9.8 V and 7.3 V, respectively. The spectra show continuous tuning of excitonic features with piezo voltage. The energy of PL peaks returns to the same value when the strain is cycled (Supplementary Fig. S6), indicating the efficient strain transfer and lack of slippage in the heterostructure. Fitting the piezo voltage dependent spectra reveals that all features exhibit a similar energy shift under strain (Fig. 4c and Supplementary Fig. S7). This implies the red shift of the excitonic features is from the strain effects on the band structure, in agreement with previous experimental[59,60] and theoretical reports[61].

Despite having similar strain shift rates, the change in degree of circular polarization ($\rho = \frac{\sigma^+ - \sigma^-}{\sigma^+ + \sigma^-}$), of the different exciton species is markedly different. While $\rho$ of $X^0$ and $X^-$ has little dependence on the strain, $X^{-\prime}$ decreases about 30% (from 0.25 to 0.18) as strain increases (Fig. 4d). The measured decrease in circular polarization with increasing strain has been observed in previous works with explanations including the decrease in the energy difference of the valence band at the $\Gamma$ and K points of the Brillouin zone[62], a reduction of the 3-fold rotational symmetry due to the uniaxial strain[63], and changes in the electron-hole exchange interaction[64]. Modeling the contribution of these different effects to our experimental observations is an interesting direction for future theoretical investigations now that we present a technique that enables the independent tuning of uniaxial strain and carrier density on TMD heterostructures. Understanding the origins of the strain-dependent $\rho$ may provide useful information for clarifying the origin of the $X^{-\prime}$ peak since it has a larger strain dependence than the other exciton and trion features. In this regard, the expanded strain range combined with gating capabilities in our devices may provide a powerful tool for both controlling and understanding excitonic physics in 2D semiconductors.

In conclusion, we have explored the strain transmission characteristics in vdW crystals and heterostructures. We found that recently studied orthorhombic crystals CrSBr, α-$MoO_3$, and $Bi_2SeO_5$ are capable of sustaining strains exceeding 1%, even in the absence of metal clamping layers. Moreover, the strain is efficiently transferred to other materials stacked on top in vdW heterostructures. These properties enable the fabrication of high-quality straintronic vdW devices capable of sustaining strains well over 1% at cryogenic temperatures. Moreover, our work provides new guidance and opportunities for engineering strain transmission in vdW devices. For instance, fabricating heterostructures which combine materials with weak and strong strain transmission, i.e., graphite and $Bi_2SeO_5$, respectively, may enable the design of on-demand and in-situ tunable strain gradients. These results establish the fundamentals for the combined exploration of electrostatic gating and large homo- and heterogeneous strains effects on 2D quantum matter.

**Supporting Information:** Additional experimental details and methods, including a schematic of the experimental setup, further Raman measurements and optical images of strained thin bulk CrSBr and CrSBr/hBN heterostructure, and additional strain and gate dependent photoluminescence measurements of the monolayer $WS_2$ / $Bi_2SeO_5$ heterostructure.

**Acknowledgements:** The strain controlled optical measurement is mainly supported by DE-SC0018171. The strain device engineering is partially supported by AFOSR FA9550-19-1-0390 and FA9550-21-1-0460. CrSBr crystal synthesis is supported by the Center on Programmable Quantum Materials, an Energy Frontier Research Center funded by the U.S. Department of Energy (DOE), Office of Science, Basic Energy Sciences (BES), under award DE-SC0019443. DGC is supported by the Columbia MRSEC on Precision-Assembled Quantum Materials (PAQM) (DMR-2011738). K.W. and T.T. acknowledge support from the JSPS KAKENHI (Grant Numbers 21H05233 and 23H02052) , the CREST (JPMJCR24A5), JST and World Premier International Research Center Initiative (WPI), MEXT, Japan. XX acknowledges support from the State of Washington funded Clean Energy Institute and from the Boeing Distinguished Professorship in Physics. ZL and JHC acknowledge the support of the David and Lucile Packard Foundation

Figures:

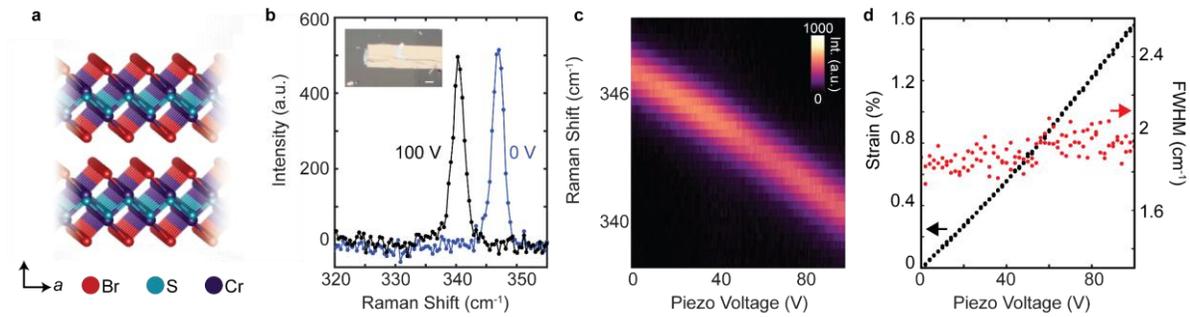

**Figure 1 | Efficient interlayer strain transmission in van der Waals crystal CrSBr. a,** Side view of the CrSBr crystal structure, consisting of monolayer planes stacked along the *c* axis. **b,** Raman spectra taken on a thick (> 100 nm, optical image inset) exfoliated CrSBr flake with 0 V (blue) and 100 V (black) applied to the piezoelectric strain cell. The phonon mode shifts by ~ 6.8 cm$^{-1}$, corresponding to a strain of ~ 1.6 % (see Methods). Scale bar 10 µm. **c**, Intensity color plot of the Raman spectra as the piezo voltage as a function of piezo voltage applied to the strain cell. **d**, Measured strain determined from the piezo voltage dependent Raman spectra in **c**. The thick flake shows a continuous shift with little change to the full-width at half maximum (FWHM) of the Raman peak (red).

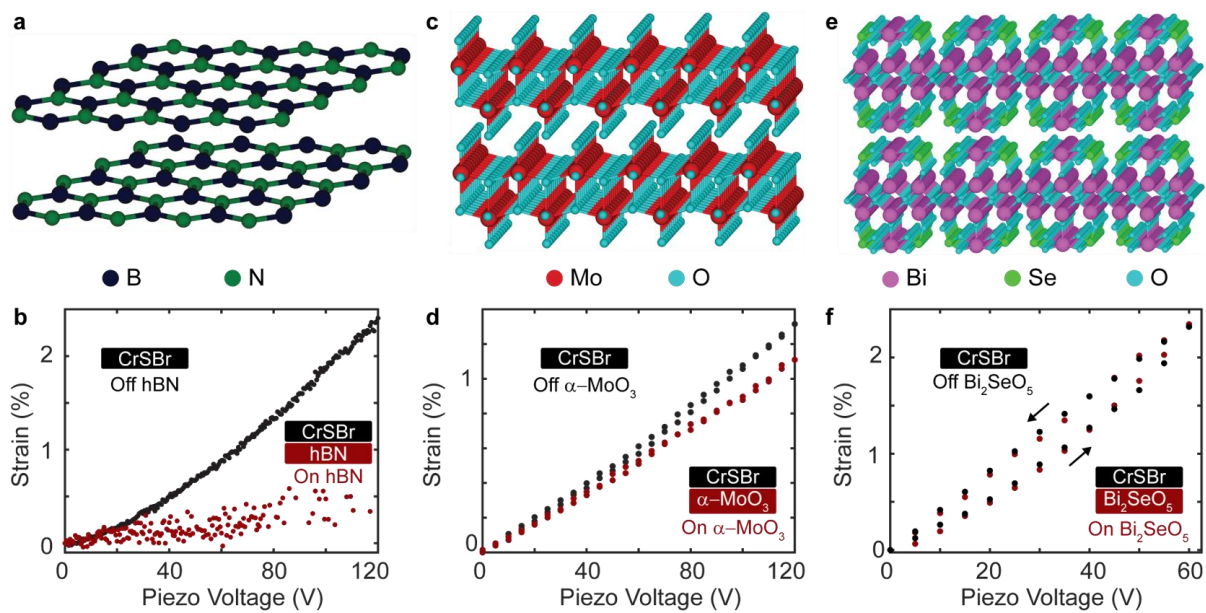

**Figure 2 | Measuring strain transmission in vdW crystals via a CrSBr strain sensor. a,** Crystal structure of hexagonal boron nitride (hBN). **b,** Measured strain of a CrSBr strain sensor placed directly on the polyimide substrate (black) and through a thin bulk (~20-30 nm) hBN crystal as the piezo voltage is increased and then decreased. **c-f,** Side view crystal structure and strain transmission characteristics of α-MoO$_3$ (**c-d**) and Bi$_2$SeO$_5$ (**e-f**). The strain in all measurements is determined by using the calibrated Raman response of a CrSBr flake deposited on top as a sensor with the zero piezo voltage value treated as zero strain (see Methods). After piezo voltage cycling (indicated by the arrows), the strain returns to the starting value, indicating negligible slippage.

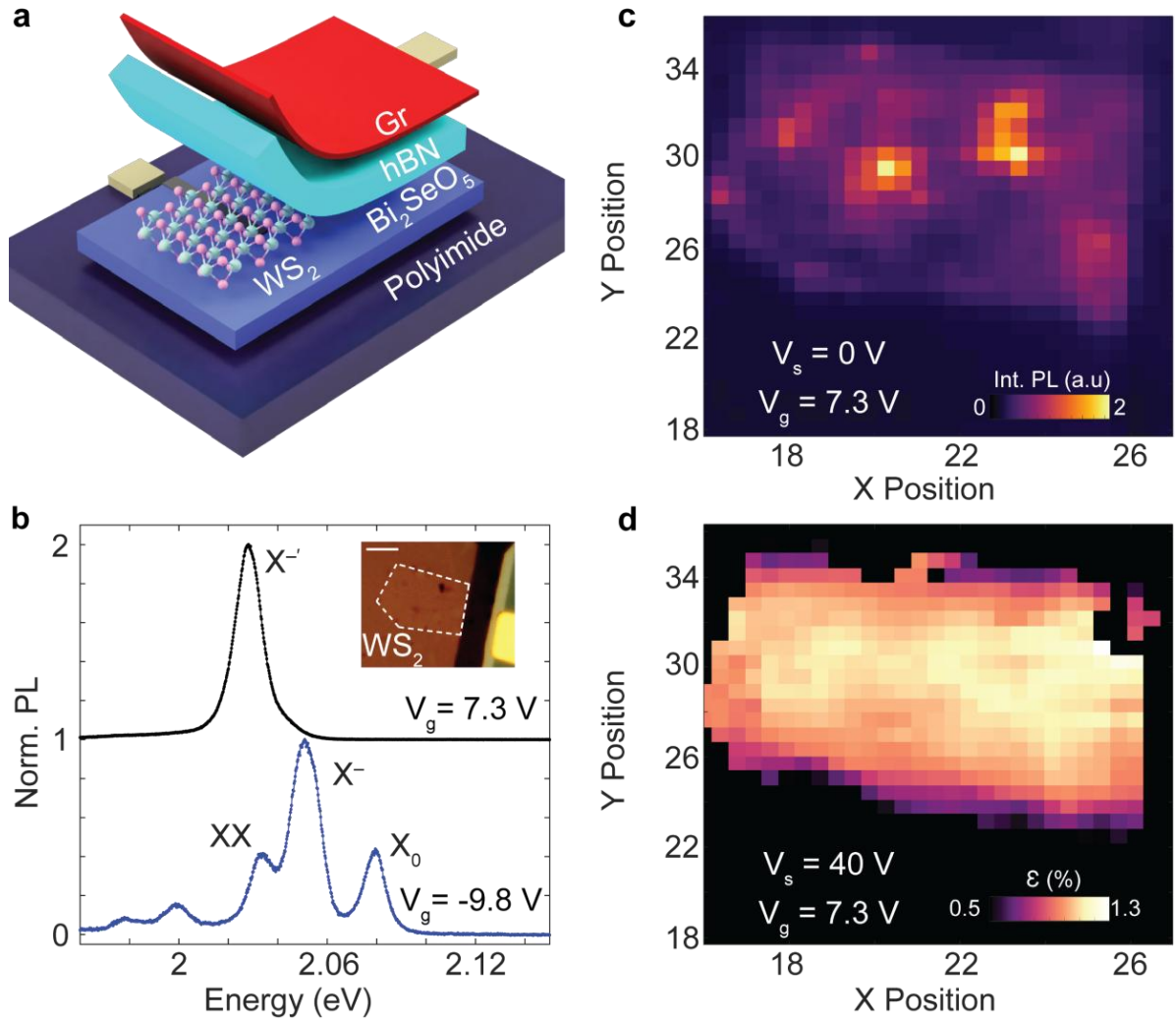

**Figure 3 | Bi$_2$SeO$_5$: a promising straintronic dielectric substrate. a,** Schematic of top-gated strain tuned vdW device. A standard graphite/hBN top gate is used to pick up and stack a monolayer WS$_2$ flake onto a pre-patterned platinum contact (grey) deposited on top of the bottom Bi$_2$SeO$_5$ layer. The entire device is assembled on a flexible polyimide substrate which is then attached to the strain cell. **b,** Photoluminescence measurements of the monolayer WS$_2$ device at top gate voltages V$_g$ of -9.8 V (blue) and 7.3 V (black). An optical image of the device is inset, scale bar 5 µm. **c,** Spatial map of PL intensity integrated over the entire spectral range with V$_g$=7.3 V. The strain cell voltage V$_s$ is set at zero. **d,** Spatial mapping of tensile strain taken at V$_s$=40 V using WS$_2$ PL (V$_g$=7.3 V) as a sensor (see Methods).

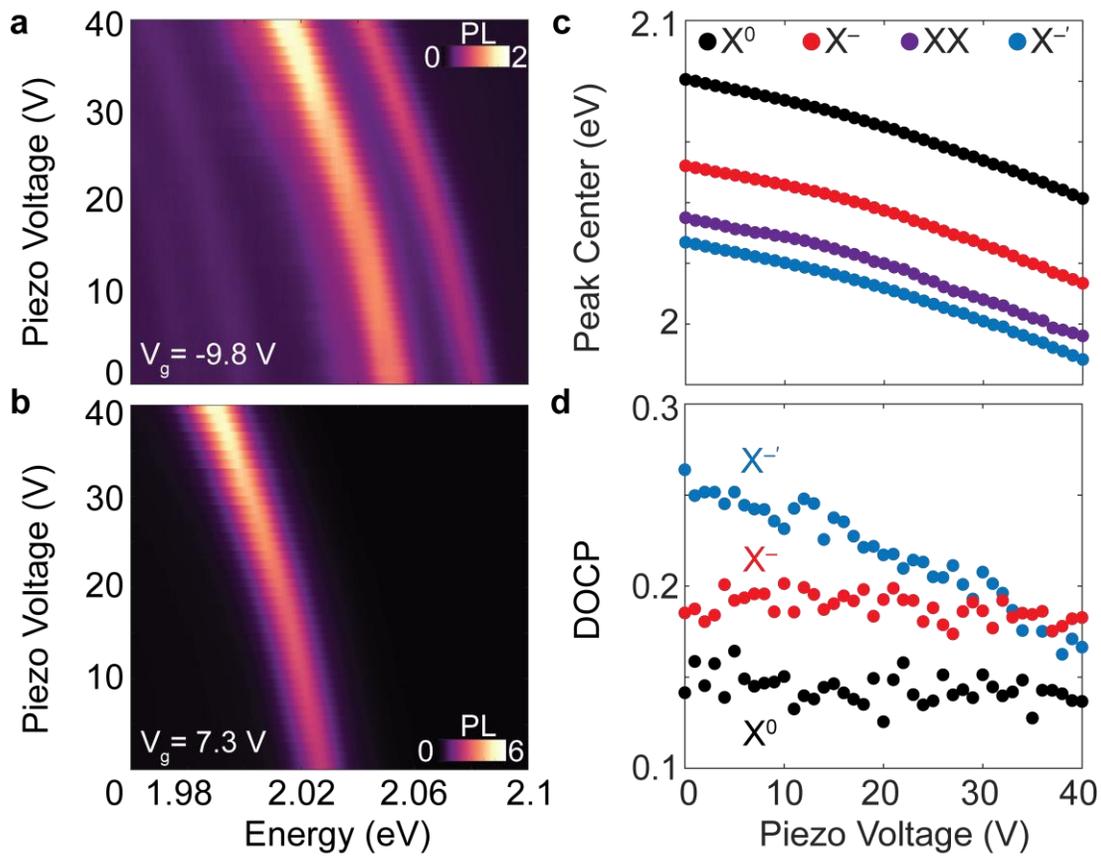

**Figure 4 | Strain tuning of exciton species in monolayer WS$_2$. a,b,** Strain-dependent co-circularly polarized PL with $V_g$ = -9.8 V (**a**) and 7.3 V (**b**). **c,** Energy and degree of circular polarization, DOCP = $\frac{\sigma^+ - \sigma^-}{\sigma^+ + \sigma^-}$, of the various PL peaks as a function of piezo voltage obtained from the spectra in **a** and **b**.

**For Table of Contents Only:**

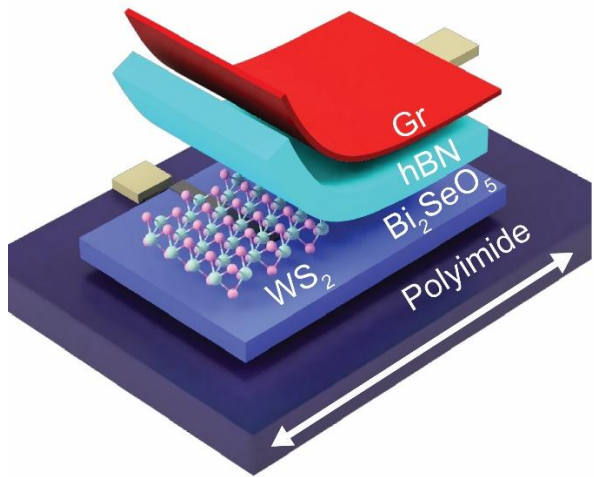 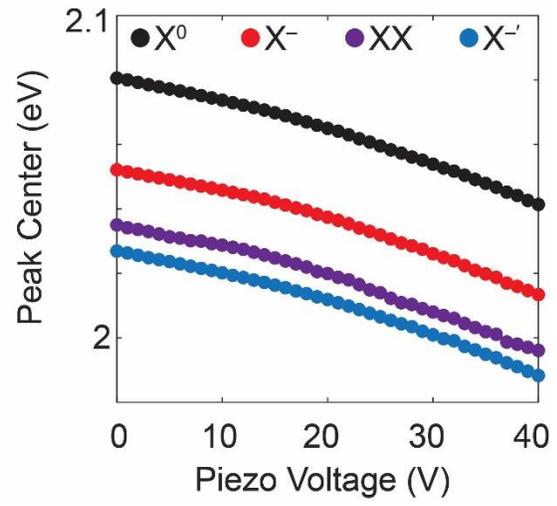

# Supplementary Information for

# Engineering robust strain transmission in van der Waals heterostructure devices

**Authors:** John Cenker[1,†*,a], Jordan Fonseca[1,†,b], Mai Nguyen[1,†], Chaowei Hu[1], Daniel G. Chica[2], Takashi Taniguchi[3], Kenji Watanabe[4], Xiaoyang Zhu[2], Xavier Roy[2], Jiun-Haw Chu[1], Xiaodong Xu[1,5,*]

[1]Department of Physics, University of Washington, Seattle, Washington 98195, USA
[2]Department of Chemistry, Columbia University, New York, NY 10027 USA
[3]Research Center for Materials Nanoarchitectonics, National Institute for Materials Science, 1-1 Namiki, Tsukuba 305-0044, Japan
[4]Research Center for Electronic and Optical Materials, National Institute for Materials Science, 1-1 Namiki, Tsukuba 305-0044, Japan
[5]Department of Materials Science and Engineering, University of Washington, Seattle, Washington 98195, USA
Current Affiliation: [a] Department of Physics, Columbia University, New York, NY, 10027, USA
[b] National Institute of Standards and Technology, Quantum Sensors Division, Boulder, CO 80305, USA

*Corresponding author's email: jc6267@columbia.edu, xuxd@uw.edu

**List of content:**

Methods

Figure S1 | Strain cell design and operation

Figure S2 | High strain behavior in unclamped thin bulk CrSBr

Figure S3 | Optical images of CrSBr/hBN heterostructure before and after applying strain.

Figure S4 | Gate-dependent photoluminescence of monolayer $WS_2$ strain device

Figure S5 | Spatial map of photoluminescence energy at low and high strains.

Figure S6 | Photoluminescence measurements taken during strain cycling.

Figure S7 | Change in energy of exciton features with respect to neutral exciton under strain.

## Methods

### Sample fabrication and strain calibration

Bulk CrSBr crystals were grown following the same recipe[1] as before, while the bulk $Bi_2SeO_5$ crystal was grown following the recipe in Ref. [2]. The bulk α-$MoO_3$ and $WS_2$ crystals were purchased from a commercial source (2D Semiconductors).

The bulk crystals were exfoliated using standard methods and suitable flakes were identified by optical contrast. The strain sensing heterostructures were then assembled through a dry transfer technique with a stamp consisting of a polypropylene carbonate (PPC) film placed on a polydimethylsiloxane (PDMS) cylinder. The CrSBr strain sensor flake was picked up followed by the target crystal (either hBN, α-$MoO_3$, or $Bi_2SeO_5$) and deposited onto the polyimide strain substrate. The strain substrate consisted of transparent 20 µm thick polyimide epoxied onto flexure sample plates produced by Razorbill instruments with Stycast 2850 FT epoxy. The long axis of the CrSBr flake was aligned with the strain axis for consistency with the previous studies[3].

To fabricate the gated monolayer $WS_2$ device, we first dropped a $Bi_2SeO_5$ flake on the polyimide substrate and evaporated platinum contacts which were connected to larger Cr/Au pads following standard lithography techniques. After patterning the $Bi_2SeO_5$, we cleaned the surface with several rounds of contact mode AFM cleaning. Then, we picked up graphite, hBN, and monolayer $WS_2$ with a polycarbonate (PC) / PDMS stamp. During stacking, we aligned the long dimension of the flakes with the strain axis but did not precisely control the relative orientation between the crystal axes of the subsequent layers. However, given the large lattice mismatch between the constituent flakes of the heterostructure, we do not expect any impact from moiré superlattices on the strain transmission as has been reported previously in small twist-angle TMDC heterostructures[4]. The device was completed by melting this stack down so that the monolayer $WS_2$ contacted the platinum, and the top gate graphite contacted a separate gold contact.

After fabrication, the sample plate was screwed into a symmetric three-piezo strain cell with the same design used in our previous experiments[4]. For the strain sensing measurement, we used the same Raman shift rate of ~ 4.2 $cm^{-1}$/% for the mode near ~ 345 $cm^{-1}$ that we determined previously[4]. In all measurements, we use the zero piezo voltage value as the zero-strain reference, which does not account for potential built-in strains during fabrication. Additionally, we note that there are a variety of reported PL shift rates for TMDC monolayers[5], likely due to the inconsistent and inefficient strain transfer between the substrate and monolayer. For this study, we used a rate of 50 meV/%, which is consistent with the majority of the literature values and the theoretical value of 56 meV/% obtained from GW-BSE calculations[6]. The energies of the optical features were obtained by fitting the Raman and PL spectra with Lorentzian and Gaussian functions, respectively.

### Optical measurements:

Optical measurements were performed using a backscattering geometry in two closed-cycle helium cryostats: Fig. 1 and Fig. 2b by Opticool from Quantum Design and all other measurements from a Montana Instruments cryostat. For Raman measurements, we used He/Ne laser (632.8 nm) with a spot size of ~ 1 µm with the power below 1mW. The collected signal was dispersed using a 1800 $mm^{-1}$ groove-density grating and then detected by an LN-cooled charge-coupled device (CCD). BragGrate$^{TM}$ notch filters were used to filter out Rayleigh scattering down to ~10 $cm^{-1}$. A roughly linear background from the polyimide substrate was subtracted to increase the accuracy of the fitting results. For photoluminescence measurements of $WS_2$, we used a laser wavelength of 532 nm and power of 10 µW. The collected PL was dispersed by a 600 mm$^{-1}$ groove-density grating and detected by a CCD.

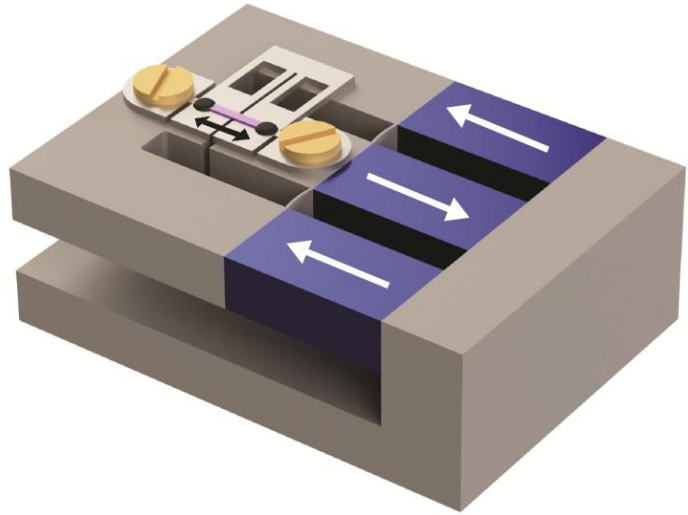

**Figure S1 | Strain cell design and operation.** A schematic of the symmetric piezoelectric strain cell used to apply uniaxial strain to the sample. When a positive piezo voltage is applied to the piezo stacks (blue), the outer stacks lengthen while the inner one contracts (indicated by white arrows). This causes the gap to expand (black arrow), applying a uniaxial tensile strain to the sample (purple).

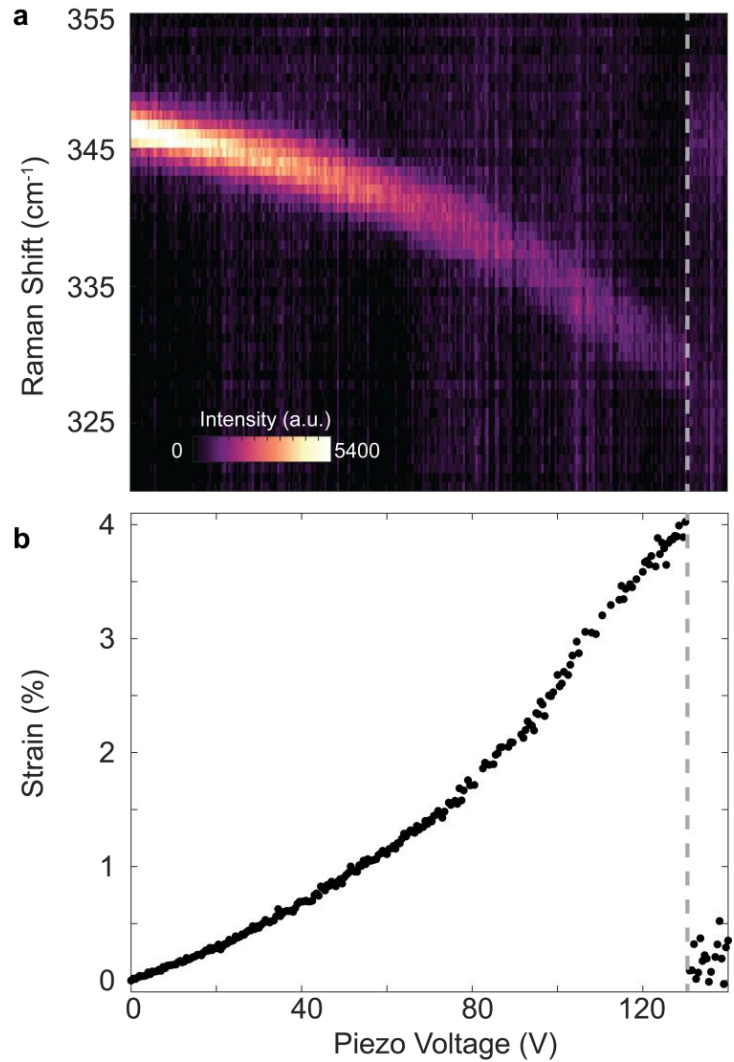

**Figure S2 | High strain behavior in unclamped thin bulk CrSBr. a,** Colormap of Raman scattering from the $P_3$ peak as a function of piezo voltage. **b,** Extracted peak energy obtained by fitting the spectra with Lorentzian functions. At a large strain approaching 4% (denoted by the grey line), the sample suddenly slips and the strain is released. The measurement was performed at a nominal temperature of 40 K to increase the piezo displacement and applied strain.

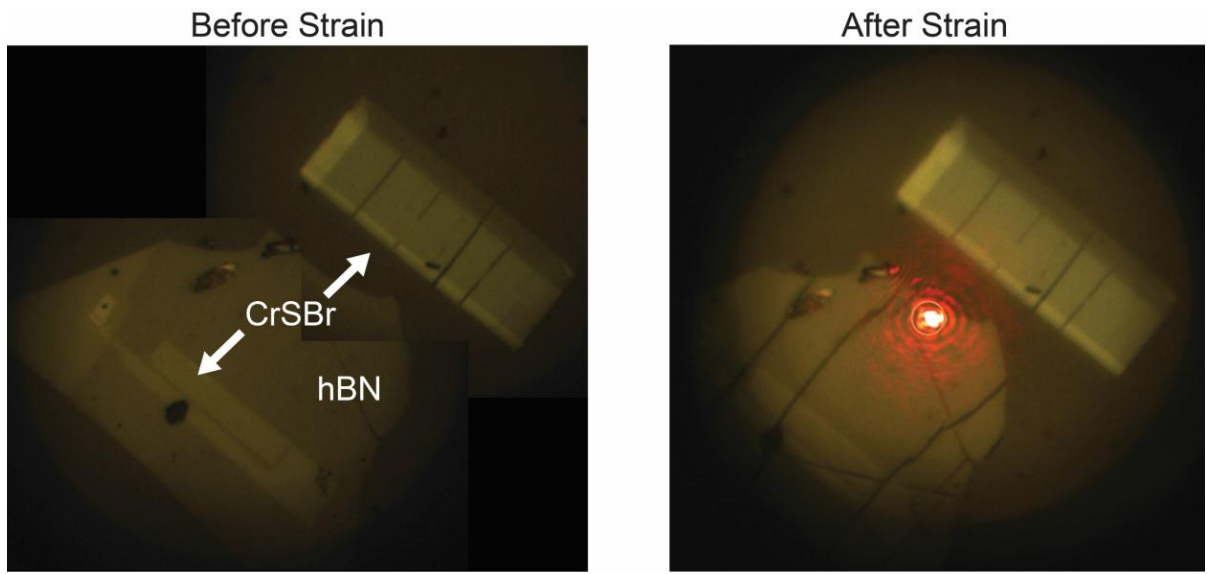

**Figure S3 | Optical images of CrSBr/hBN heterostructure before and after applying strain.** Optical image of a thin bulk CrSBr flake and a CrSBr / hBN heterostructure deposited on the same polyimide substrate at 0 V before (left) and after (right) a piezo voltage of 120 V was applied to the strain cell. The hBN heterostructure exhibits a variety of new wrinkles after strain cycling, while the CrSBr flake does not.

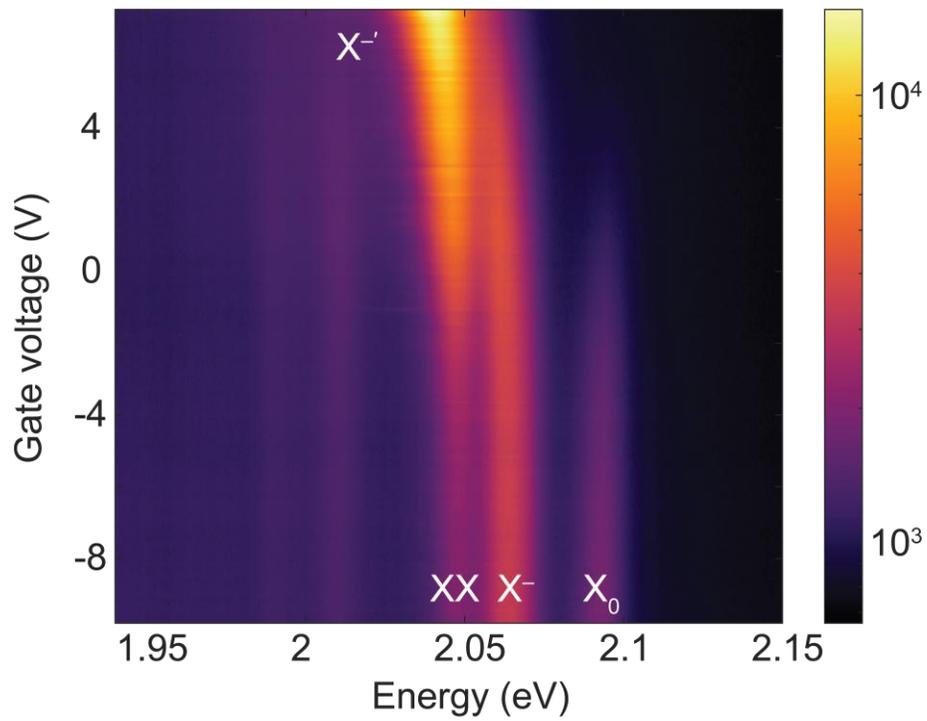

**Figure S4 | Gate-dependent photoluminescence of monolayer WS$_2$ strain device.** PL measurements as a function of top gate voltage. The piezo voltage to strain cell is 0 V. The sample is excited by $\sigma^+$ light, and the detected signal is co-circularly polarized with respect to the excitation.

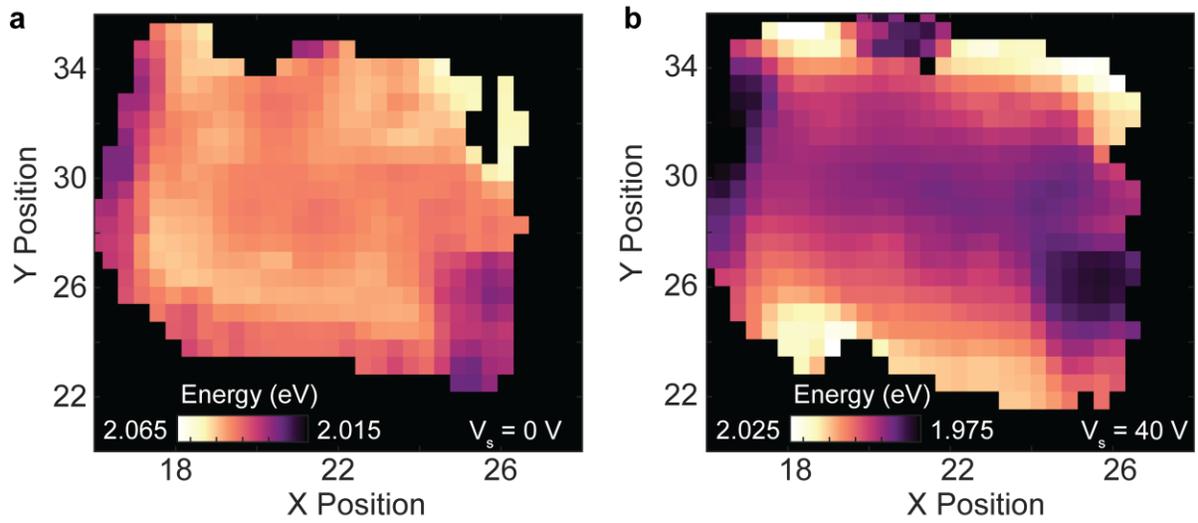

**Figure S5 | Spatial map of photoluminescence energy at low and high strains.** Spatial maps of PL peak energy at strain voltages of 0 V (**a**) and 40 V (**b**) with the top gate voltage $V_g = 7.3$ V. The peak center energy is obtained by a Gaussian fit of $X^{-}$ at each coordinate.

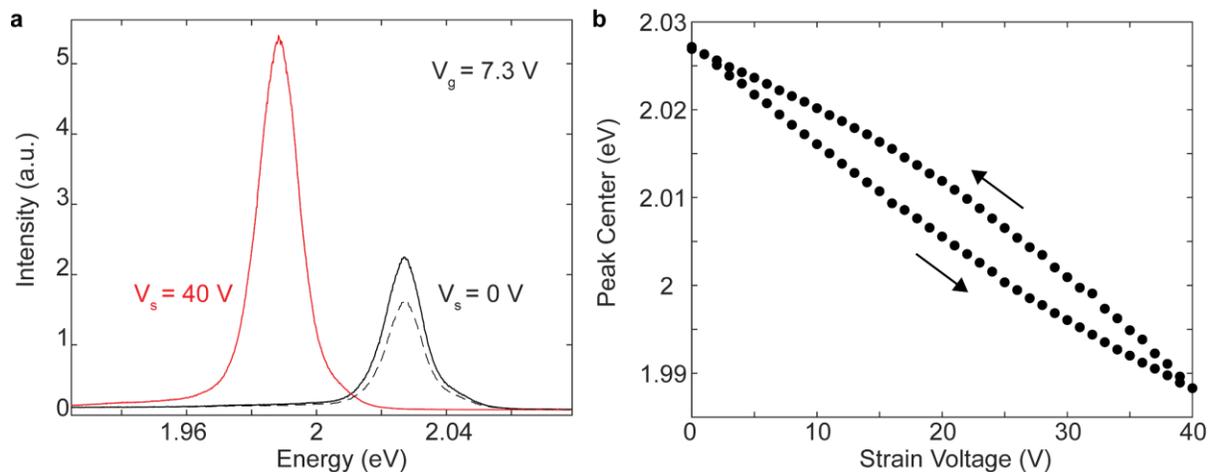

**Figure S6 | Photoluminescence measurements taken during strain cycling. a,** PL spectra taken as the strain is swept from 0 V (solid black curve) to 40 V (red) and then back to 0 V (dashed black curve). The gate voltage is kept fixed at 7.3 V.

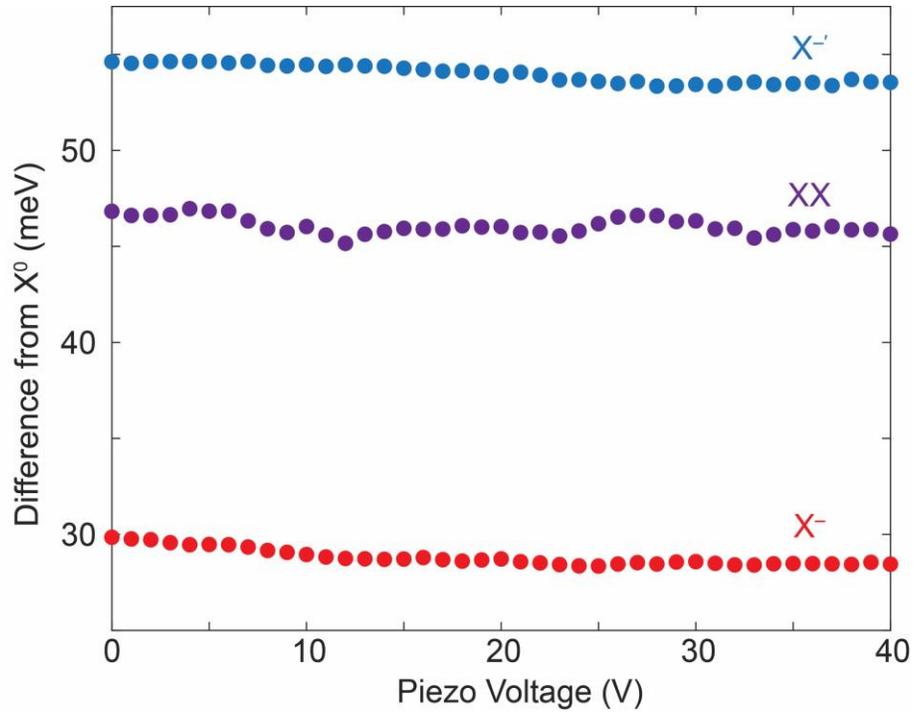

**Figure S7 | Change in energy of exciton features with respect to neutral exciton under strain.** The difference in energy between the neutral exciton ($X^0$) and the different exciton features as a function of piezo voltage. The energy spacing between the exciton species shows negligible change (~ 1 meV) in comparison with the overall strain-induced redshift of the features (~ 40 meV, see Fig. 4c). Such subtle changes may arise due to small differences in the redshift rates, or due to extrinsic factors such as shifting of the beam spot position during the strain sweep.